\documentclass[10pt,twocolumn,tightenlines,floats,aps,amsmath,amssymb,prd,floatfix,longbibliography,superscriptaddress,notitlepage,nofootinbib]{revtex4-2}

\usepackage[utf8]{inputenc}
\usepackage{wasysym}
\usepackage{amsmath,amssymb,amsfonts,amsthm,mathrsfs}
\usepackage{graphicx,wrapfig}
\usepackage{enumerate}
\usepackage{color}

\usepackage{physics}
\usepackage{mathtools}
\usepackage{dcolumn}
\usepackage{orcidlink}

\usepackage{hyperref}
\hypersetup{
	bookmarks=true,         % show bookmarks bar?
	unicode=false,          % non-Latin characters in Acrobat’s bookmarks
	pdftoolbar=true,        % show Acroba,t’s toolbar?
	pdfmenubar=true,        % show Acrobat’s menu?
	pdffitwindow=false,     % window fit to page when opened
	pdfstartview={FitH},    % fits the width of the page to the window
	pdftitle={Squeezed vacua and primordial features in effective theories of inflation at N2LO},    % title
	pdfauthor={Mauricio Gamonal, Eugenio Bianchi},     % author
	pdfkeywords={inflation, cosmology, quantum gravity}, % list of keywords
	pdfnewwindow=true,      % links in new PDF window
	colorlinks=true,       % false: boxed links; true: colored links
	linkcolor=red,          % color of internal links (change box color with linkbordercolor)
	citecolor=blue,        % color of links to bibliography
	filecolor=cyan,         % color of file links
	urlcolor=blue        % color of external links
}

\newcommand{\ii}{\mathrm{i}}
\newcommand{\ee}{\mathrm{e}}

\newcommand{\outcomment}[1]{}

\newcommand{\comment}[1]{#1}

\begin{document}

\title{Squeezed Vacua and Primordial Features in Effective Theories of Inflation at N2LO}

\author{Eugenio Bianchi\,\orcidlink{0000-0001-7847-9929}}
\thanks{}
\email{ebianchi@psu.edu}
\author{Mauricio Gamonal\,\orcidlink{0000-0002-0677-4926}}
\thanks{}
\email{mgamonal@psu.edu}

\affiliation{Institute for Gravitation and the Cosmos, The Pennsylvania State University, University Park, Pennsylvania 16802, USA}\affiliation{Department of Physics, The Pennsylvania State University, University Park, Pennsylvania 16802, USA}

\date{\today}

\begin{abstract}
A finite duration of cosmic inflation can result in features $\mathcal{P}_{\mathcal{R}}(k) = |\alpha_k-\beta_k\,\mathrm{e}^{\mathrm{i}\delta_k}|^2 \,\mathcal{P}_{\mathcal{R}}^{(0)}(k)$ in the primordial power spectrum that carry information about a quantum gravity phase before inflation. While the almost scale-invariant power spectrum $\mathcal{P}_{\mathcal{R}}^{(0)}$ for the quasi-Bunch-Davies vacuum is fully determined by the inflationary background dynamics, the Bogoliubov coefficients $\alpha_k$ and $\beta_k$ for the squeezed vacuum depend on new physics beyond inflation and have been used to produce phenomenological templates for the features. The phase $\delta_k$ vanishes in de Sitter space and therefore is often neglected, but it results in non-trivial effects in quasi-de Sitter inflationary geometries. Here we consider a large class of effective theories of inflation and provide a closed-form expression for $\delta_k$ and for the fully expanded power spectrum up to next-to-next-leading order (N2LO) in the Hubble-flow expansion. In particular, for the Starobinsky model of inflation we find that this relative phase can be expressed in terms of the scalar tilt $n_\mathrm{s}$ as $\delta_{k\ast}=\frac{\pi}{2}(n_\mathrm{s}-1)-\frac{\pi}{4}(n_\mathrm{s}-1)^2\,\ln(k/k_*)$. The relative phase results in a negative shift and a running frequency that have been considered in the most studied phenomenological templates for primordial features, thus providing precise theoretical predictions for upcoming cosmological observations.
\end{abstract}

\maketitle
% Add the custom footnote with a double dagger marker
\begingroup
\renewcommand{\thefootnote}{$\ddagger$} % Use double dagger in math mode
\footnotetext{Both authors contributed equally to this work.}
\endgroup

% Reset footnote counter to ensure text footnotes start at 1
\setcounter{footnote}{0}

\section{Introduction}
\label{sec:Introduction}

Cosmic inflation \cite{Brout1978, Starobinsky1979, Starobinsky1980, Sato:1981ds,Sato:1981qmu,Guth1981, Mukhanov1981,Linde1982,Albrecht1982,Guth1982,Hawking1982,Linde1983}, a phase of quasi-de Sitter expansion in the early Universe with quantum perturbations initially in the vacuum state \cite{Mukhanov1992,Martin:2024qnn}, predicts a prototypical power spectrum of primordial curvature perturbations of the form $\mathcal{P}^{(0)}_{\mathcal{R}}(k) \approx   (k/k_\ast)^{n_{\mathrm{s}}-1}\, \mathcal{A}_{\mathrm{s}}$ that is nearly scale-invariant and strongly constrained by cosmic microwave background (CMB) observations \cite{Calabrese:2013jyk,Planck:2018jri,ACT:2020frw}. As we enter an era of precision cosmology, upcoming experiments \cite{CORECollaboration2016,SimonsObservatory:2018koc,S4Collaboration2020, LiteBIRDCollaboration2022,Euclid:2023shr,Mergulhao:2023ukp,Antony:2024vrx} have been designed to probe primordial features in the curvature power spectrum that go beyond the standard inflationary paradigm, which are usually parametrized by phenomenological templates of the form 
\begin{equation}
\mathcal{P}^{(\mathrm{phen})}_{\mathcal{R}}(k)=\Big[1-R_k\cos\big(\Xi_k+\delta\big)\Big]\,\left(\frac{k}{k_\ast}\right)^{n_\mathrm{s}-1}\mathcal{A}_s,
\label{eq:template-phase}
\end{equation}
allowing power suppression and oscillations \cite{Achucarro:2010da,Hazra:2010ve,Chen:2011zf,Jackson:2013vka,Flauger:2014ana,Ballardini:2016hpi,Palma:2017wxu,LHuillier:2017lgm,Zeng:2018ufm,Ballardini:2018noo,Beutler:2019ojk,Domenech:2019cyh,Slosar:2019gvt,Braglia:2020taf,Hamann:2021eyw}. However, further theoretical inputs are required to properly constrain the functions $R_k$ and $\Xi_k$ with observations. 

A specific choice of squeezed vacuum for cosmological perturbations, motivated by new physics in a pre-inflationary phase 
\cite{Starobinsky:1992ts,Niemeyer:2000eh,Tanaka:2000jw,Starobinsky:2001kn,Hui:2001ce,Goldstein:2002fc,Danielsson:2002kx,Niemeyer:2002kh,Contaldi:2003zv,Martin:2003kp,Armendariz-Picon:2003knj,Bozza:2003pr,Danielsson:2004xw,Wang:2007ws,Destri:2009hn,Lello:2013mfa,Lello:2013awa,Chen:2015gla,Broy:2016zik,Handley:2016ods,Kaloper:2018zgi,Gessey-Jones:2021yky,Letey:2022hdp,Schwarz:2009sj,Ramirez:2011kk,Ramirez:2012gt,Cicoli:2014bja,CastelloGomar:2017kbo,Bhardwaj:2018omt,Kowalczyk:2024ech,MenaMarugan:2024zcv,Akama:2024vgu}, results in a squeezed power spectrum of the form $   \mathcal{P}_{\mathcal{R}}(k)=\big|\alpha_k-\beta_k\,\mathrm{e}^{\mathrm{i}\delta_k}\big|^2\;\, \mathcal{P}_{\mathcal{R}}^{(0)}(k)$, that can provide a top down derivation of the phenomenological template \eqref{eq:template-phase} \cite{Chluba:2015bqa}. The Bogoliubov coefficients $\alpha_k$ and $\beta_k$ are determined by the choice of vacuum state \cite{Birrell:1982ix,Bunch:1978yq} and carry information about the quantum gravity phase that preceded inflation, e.g., as described by loop quantum cosmology \cite{Ashtekar:2011ni,Agullo:2023rqq,Ashtekar2021}, which provides specific prescriptions for the pre-inflationary regime \cite{Agullo:2012sh,Fernandez-Mendez:2012poe,Fernandez-Mendez:2013jqa,Agullo:2013ai,Agullo:2015tca,deBlas:2016puz,Ashtekar:2020gec,Navascues:2021mxq,Martin-Benito:2023nky,ElizagaNavascues:2023xah,MenaMarugan:2024vyy,Garay:2024afl,Montese:2024ypz}. 

The relative phase $\delta_k$ vanishes in exact de Sitter space and therefore is often ignored. In this paper, we compute the phase $\delta_k$ in quasi-de Sitter space and study its effect on the primordial power spectrum for a wide family of effective theories of inflation, at next-to-next-to-leading order (N2LO) in a squeezed vacuum state, by using the Green's function method \cite{Stewart2001,Auclair2022,Ballardini:2024irx}. In particular, we show that the phase $\delta_k$ is, in principle, observable and fully determined by the inflationary background dynamics, providing a closed expression in terms of Hubble-flow parameters \cite{Schwarz:2001vv}, up to N2LO. This allows us to make precise theoretical predictions for upcoming cosmological observations, which can further constrain the phenomenological templates \eqref{eq:template-phase}.

\section{Effective theories of inflation}
\label{sec:EFTs}
We adopt the approach developed in \cite{Bianchi:2024qyp} for cosmological perturbations in effective theories of inflation. We start from a classical background spacetime described by a spatially flat Friedmann–Lema\^itre–Robertson–Walker (FLRW) metric, with scale factor $a(t)$. Small perturbations of the geometry and matter fields induce physical degrees of freedom that are encoded in scalar-vector-tensor (SVT) modes $\Psi(\vb{x},t)$ \cite{Mukhanov1992}. In its most general form \cite{Bianchi:2024qyp}, the quadratic contributions to the action for each SVT mode reads 
% \begin{align}
% 	\label{eq:Quadratic-Action-psi}
% 	S_{2}[\psi] &=  \int \dd{t} \int \frac{ \dd[3]{\vb{k}} }{(2\pi)^3}\, a(t)^3 Z_\psi(t) \Bigg[  \frac{1}{2}\big|\dot{\psi}(\vb{k},t)\big|^2 \nonumber  \\
%  &\hspace{7em} -\frac{1}{2} c_\psi(t)^2\,  \frac{k^2}{a(t)^2} \abs{\psi(\vb{k},t)}^2 \Bigg],
% \end{align}
\begin{equation}
	\label{eq:Quadratic-Action-psi}
 S= \! \int\!\! \dd{t} \!\!\int\!\! \tfrac{ \dd[3]{\vb{k}} }{(2\pi)^3}\, \tfrac{a(t)^3 Z_\psi(t)}{2} \Big[  \big|\dot{\psi}(\vb{k},t)\big|^2 
 -\tfrac{c_\psi(t)^2\,k^2}{a(t)^2} \abs{\psi(\vb{k},t)}^2 \Big],
\end{equation}
where $k=\abs{\vb{k}}$, and $\psi( \vb{k},t )$ is the Fourier transform of each of the SVT modes. The kinetic amplitude, $Z_\psi(t)$, and the speed of sound, $c_\psi(t)$, are two independent functions that do not depend on the mode $k$ and satisfy the conditions $Z_\psi(t)>0$, and  $c_\psi(t)^2 >0$. In a quasi-de Sitter phase, where the Hubble rate $H(t)\equiv \dot{a}(t)/a(t)$ is almost constant, we can introduce the \textit{Hubble-flow expansion} \cite{Schwarz:2001vv}, defined recursively in terms of the dimensionless parameters $\epsilon_{n\rho}(t)  \equiv - \dot{\epsilon}_{n-1\rho}(t)/\big(H(t)\epsilon_{n-1\rho}(t)\big)$, where $\epsilon_{0\rho} = \rho$, with $\rho= H$, $Z_\psi$, or $c_\psi$. For example, $\epsilon_{1H} = -\dot{H}/H^2$, $\epsilon_{2Z} = -\dot{\epsilon}_{1Z}/(H\,\epsilon_{1Z})$, etc. For a comparison of sign conventions in $\epsilon_{nH}$ we refer to the conversion table in \cite{Bianchi:2024qyp}. The Hubble-flow parameters can be understood as a measure of the deviation from exact de Sitter space ($\epsilon_{1H} = 0$), and from 
``vanilla" single-field inflation ($\epsilon_{1c}= 0$, and $\epsilon_{1Z}= \epsilon_{1H}$ for scalar modes while $\epsilon_{1Z}=0$ for tensor modes).

The primordial scalar and  tensor perturbations that describe the seeds of the large-scale structure and the CMB anisotropies are assumed to be quantum fields $\hat{\Psi} (\vb{x},t)$ initially in a  Fock vacuum $\ket{0}$, defined by $\hat{a}(\vb{k}) \ket{0} = 0\,$, $\forall\, \vb{k}$, with bosonic creation and annihilation operators, $[\hat{a}(\vb{k}),\hat{a}^\dagger(\vb{k^\prime})] = (2\pi)^3 \,\delta^{(3)} (\vb{k}-\vb{k}^\prime)$. In Fourier space, the mode expansion of the field reads $\hat{\psi} (\vb{k},t) = u(k,t) \,\hat{a}(\vb{k}) + u^\ast (k,t)\,\hat{a}^\dagger (-\vb{k})$. Generalizing the construction of the Mukhanov-Sasaki variables, we perform a time reparametrization $t\to y = -k\, \tau$, with $\tau \equiv - \tilde{c}_\psi (t)/\big( a(t) H(t)\big)$, where $\tau$ generalizes the conformal time $\tau$, and $\tilde{c}_\psi(t)$ is defined in \eqref{eq:y-in-terms-of-aH}. In parallel, the mode functions are rescaled via $u(k,t) \to \big(y\, w(y)\big)/\sqrt{2\, k^3\, \mu(y)}$, with $\mu(y) = (\hbar \, H(t)^2)^{-1}\, Z_\psi(t) \, c_\psi(t)\, \tilde{c}_\psi^2$. With these definitions, the canonical commutation relations for the quantum field result in canonical Wronskian conditions for the mode functions, i.e., $w(y)\, w'{}^\ast (y) - w'(y)\,  w^\ast (y)  = -2 \,\ii$. Moreover, the equations of motion for the field result in the mode equation
\begin{equation}
\label{eq:EOM}
    w''(y)+\qty(1-\frac{2}{y^2} )w(y) = \frac{g(y)}{y^2}\,w(y)\,,
\end{equation}
where the function $g(y) = g_{1k} + g_{2k}\, \ln(y)\,+\,\order{\epsilon^3}$ depends on the background quantities $H(t)$, $Z_\psi(t)$, and $c_\psi(t)$, with its explicit N2LO expansion given in Eq. \eqref{eq:gk}. The function $g(y)$ is slowly changing and can be expanded in a Taylor series in powers of $\ln(y)$, understood as an expansion around the peculiar time $\tau_k = -1/k$, i.e., $y_k = 1$. More details on this framework and the derivation of \eqref{eq:EOM} can be found in Appendix \ref{app:DerivationModes} and in \cite{Bianchi:2024qyp}. 

\section{Quasi Bunch-Davies vacuum}
\label{sec:Power-qBD}

In exact de Sitter space there is a distinguished vacuum state, the Bunch-Davies vacuum \cite{Bunch:1978yq,Birrell:1982ix}. It is defined by the mode function $w_{\mathrm{BD}}(y)=\left(1+\ii/y\right)\,\ee^{\,\ii y}$, which, together with its complex conjugate $w_{\mathrm{BD}}^*(y)$, provides a basis of solutions of the Wronskian condition and the mode equation $w_{\mathrm{BD}}''(y)+(1-2/y^2)\,w_{\mathrm{BD}}(y)=0$. The defining property of this basis of solutions is that the state $|\mathrm{BD}\rangle$ has correlation functions $\langle \mathrm{BD}|\hat{\Psi} (\vb{x},t) \,\hat{\Psi} (\vb{x}',t')|\mathrm{BD}\rangle$ which are ultraviolet adiabatic and respect all the symmetries of de Sitter space, not just the homogeneity and isotropy of the cosmic time slices. The order-by-order Hubble-flow expansion of $g(y)$ allows us to introduce a quasi-Bunch-Davies vacuum state $|\mathrm{qBD}\rangle$ with mode functions $w_{\mathrm{qBD}}(y)=w_{\mathrm{BD}}(y)+ w_1(y)+w_2(y)+\cdots$, defined as an expansion around the Bunch-Davies vacuum. The mode function $w_{\mathrm{qBD}}(y)$ is the unique solution of the Wronskian condition and the mode equation \eqref{eq:EOM}, determined iteratively via the Green function method \cite{Stewart2001,Auclair2022,Bianchi:2024qyp,Ballardini:2024irx}. Then, the late-time power spectrum associated to the quasi-Bunch-Davies vacuum state at N2LO is given by the expression
\begin{align}
	\label{eq:Power-BunchDavies}
&\!\!\!\mathcal{P}_{\mathrm{qBD}} (k)\equiv\lim_{t\to \infty } \frac{k^3}{2\pi^2}  |u_{\mathrm{qBD}}(k,t)|^2  = \lim_{y\to 0^+ } \frac{\abs{y\,w_{\mathrm{qBD}}(y)}^2}{4\pi^2\,\mu(y)} \nonumber \\[.5em]
&= \frac{\hbar\, H_\ast^2}{4\pi^2 Z_\ast c_\ast^3} \Bigg[p_{0\ast} + p_{1\ast} \ln(\frac{k}{k_\ast} ) + p_{2\ast} \ln(\frac{k}{k_\ast})^2 \Bigg]\,,
\end{align}
where each of the quantities is evaluated at $y_\ast = k/k_\ast$, with an associated pivot scale $k_\ast = (a_\ast H_\ast)/\tilde{c}_\ast$. Details on the derivation of this expression are briefly discussed in Appendix \ref{eq:App-B}, and in particular, the coefficients $p_{0\ast}$, $p_{1\ast}$ and $p_{2\ast}$ are given in \eqref{eq:p0}, \eqref{eq:p1} and \eqref{eq:p2}, respectively.

\section{Squeezed vacua and power spectrum}
\label{sec:Squeezed}

The quasi-Bunch-Davies vacuum $|\mathrm{qBD}\rangle$ is a natural choice of state with a huge predictive power. It can be understood as the in-vacuum for an inflationary phase that is infinitely long in the asymptotic past. Its construction, starting from causal Green functions for the mode equation \eqref{eq:EOM} expanded around the exact de Sitter background, guaranties that the state $|\mathrm{qBD}\rangle$ approaches the Bunch-Davies vacuum in the far past, together with its corrections in Hubble-flow parameters at N2LO. On the other hand, if the slow-roll inflationary phase is only transitory and is preceeded by a pre-inflationary phase, the state $|\mathrm{qBD}\rangle$ cannot be considered as the in-vacuum anymore, but it can still be used as a reference state that is completely determined by the background geometry $a(t)$, the kinetic amplitude $Z_\psi(t)$ and the speed of sound $c_\psi(t)$ for the perturbative quantum field $\hat{\Psi}(\vb{x},t)$. In particular, any pure Gaussian state with homogeneous and isotropic correlation functions can be written in terms of a two-mode squeezing of the reference state $|\mathrm{qBD}\rangle$. The mode functions $w_{\mathrm{sqz}}(y)$ that define the squeezed vacuum are related to the mode functions $w_{\mathrm{qBD}}(y)$ of the reference quasi-Bunch-Davies vacuum by the Bogoliubov transformation $w_{\mathrm{sqz}}(y)=\alpha_k\,w_{\mathrm{qBD}}(y)+\beta_k\,w_{\mathrm{qBD}}^*(y)$, with the Bogoliubov coefficients $\alpha_k$ and $\beta_k$ satisfying the canonical Wronskian condition $|\alpha_k|^2-|\beta_k|^2=1$. Equivalently, the bosonic operator defined by the Bogoliubov transformation $\hat{b}(\vb{k})=\alpha_k^*\,\hat{a}(\vb{k})-\beta^*_k \,\hat{a}^\dagger(-\vb{k})$, annihilates the state,
\begin{equation}
\!\!\!|\mathrm{sqz}\rangle=\!\frac{1}{\sqrt{\mathcal{N}}}\exp\!\Big({\!-\!\!\int\!\! \frac{d^3 \vb{k}}{(2\pi)^3} \frac{1}{2}\frac{\beta_k^*}{\alpha_k^*} \,\hat{a}^\dagger(\vb{k})\hat{a}^\dagger(-\vb{k})}\Big)|\mathrm{qBD}\rangle,
\end{equation}
defined as a two-mode squeezed state with respect to the reference $|\mathrm{qBD}\rangle$ state. The condition of ultraviolet adiabaticity imposes that the Bogoliubov coefficient $\beta_k$ approaches zero, $\beta_k\to 0$, sufficiently fast as $k\to\infty$, and the requirement that the state belongs to the Fock space built over the Fock vacuum $|\mathrm{qBD}\rangle$ imposes that the total number of excitations is finite, $\int\frac{d^3 \vb{k}}{(2\pi)^3} |\beta_k|^2<\infty$. The squeezed vacuum $|\mathrm{sqz}\rangle$ can be understood as an excited state with a finite homogeneous-and-isotropic expectation value of the energy density and pressure of  perturbations \cite{Birrell:1982ix}. In particular, the equal-time correlation function is
\begin{equation}
\langle \mathrm{sqz}|\hat{\Psi} (\vb{x},t) \hat{\Psi} (\vb{x}',t) |\mathrm{sqz} \rangle=\int_0^\infty\frac{dk}{k}\,\frac{\sin(k\,|\vb{x}-\vb{x'}|)}{k\,|\vb{x}-\vb{x'}|}\,\mathcal{P}_{\mathrm{sqz}}(k),
\end{equation}
with the power spectrum $\mathcal{P}_{\mathrm{sqz}}(k)$ for the squeezed vacuum given by
\begin{align}
\mathcal{P}_{\mathrm{sqz}}(k) &=\lim_{y\to0^+} \frac{\abs{y\,w_{\mathrm{sqz}}(y)}^2}{4\pi^2\,\mu(y)} \nonumber \\[.5em]
& =\lim_{y\to0^+} \Big|\alpha_k+\beta_k\,\frac{w_{\mathrm{qBD}}^*(y)}{w_{\mathrm{qBD}}(y)}\Big|^2\, \frac{\abs{y\,w_{\mathrm{qBD}}(y)}^2}{4\pi^2\,\mu(y)}\qquad\nonumber\\[.9em]
& =\,\big|\alpha_k-\beta_k\,\ee^{\ii\delta_k}\big|{}^2\,
\,\mathcal{P}_{\mathrm{qBD}}(k) \, ,\label{eq:power-delta}
\end{align}
with $\ee^{\ii \delta_k}=-\lim_{y\to0^+} w_{\mathrm{qBD}}^*(y)/w_{\mathrm{qBD}}(y)$, which is completely determined by the qBD mode functions. This expression can be computed order-by-order in a Hubble-flow expansion using the asymptotic relations \eqref{eq:limits1}, \eqref{eq:limits2}, \eqref{eq:limits3}. %There is a highly nontrivial check now: for the limit $y\to 0^+$ to be finite, all the $\ln(y)$ terms have to cancel exactly, which corresponds to the freezing  limit of the power spectrum. 
In particular, in the limit of exact de Sitter background, i.e., at the leading order (LO), the BD mode functions are purely imaginary, causing the phase $\delta_k$ to vanish. However, at NLO and at N2LO, the phase receives nontrivial contributions. Specifically,  at N2LO we find
\begin{equation}
\delta_k =-\frac{\pi}{3}g_{1k}\,+\,\frac{\pi}{27}\big(g_{1k}^2+(9\,C-3)\,g_{2k}\big)\;+\;\order{\epsilon^3}\,,
\end{equation}
where $C=\gamma_E+\ln(2)-2\simeq -0.730$, and the exact expressions for $g_{1k}$ and $g_{2k}$ are found in Eq. \eqref{eq:gk}.~It is useful to parametrize the Bogoliubov coefficients as $\alpha_k=\;\cosh(r_k)\,\ee^{\ii\theta_k}$ and  $\beta_k=\;\sinh(r_k)\,\ee^{\ii(\theta_k+\phi_k)}$, with the parameter $r_k\geq 0$ controlling the amount of squeezing, an overall unobservable phase $\theta_k$, and a physical relative phase $\phi_k\in [0,2\pi)$. In terms of these parameters, we can define a squeezing factor $\Upsilon_{\mathrm{sqz}}(k) = \mathcal{P}_{\mathrm{sqz}(k)}/\mathcal{P}_{\mathrm{qBD}}(k) = \cosh(2r_k)-\sinh(2r_k)\, \cos(\phi_k+\delta_k)$. Note that, while the relative phase $\phi_k$ depends on the relation between the two states $|\mathrm{qBD}\rangle$ and $|\mathrm{sqz}\rangle$, the phase $\delta_k$ is purely determined by the Hubble-flow parameters of the background.

\begin{figure*}[t]
	\includegraphics[width=0.49\linewidth]{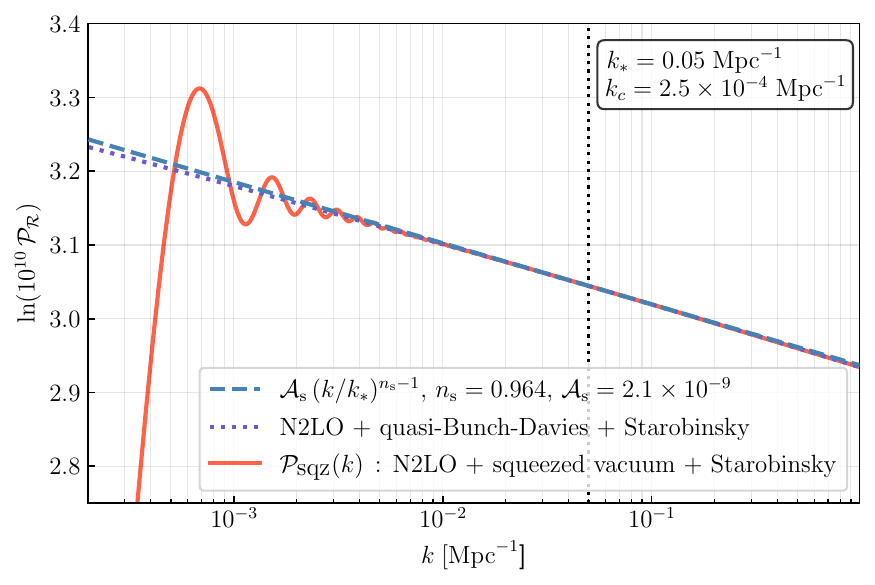}\quad
	\includegraphics[width=0.49\linewidth]{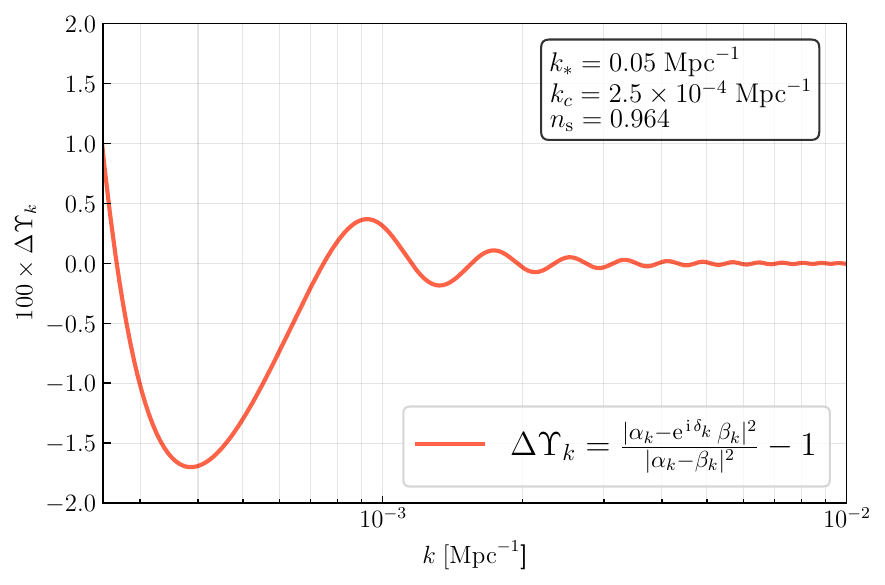}
	\caption{(Left): As an illustrative example, we compare the squeezed power spectrum of curvature perturbations $\mathcal{P}_{\mathrm{sqz}}(k)$ (solid red line) corresponding to the Bogoliubov coefficients \eqref{eq:alpha-beta-Bogoliubov}, with an exact power-law (dashed blue) and the N2LO expression for the qBD state \eqref{eq:Power-BunchDavies} (dotted black). The values of the Hubble-flow parameters are the ones predicted by Starobinsky inflation for a fiducial value $N_\ast = 55$, as discussed in \cite{Bianchi:2024qyp}. The vertical black dotted line corresponds to the pivot scale $k_\ast$ typically used in CMB observations. (Right): \comment{
    We illustrate the effect of the phase $\delta_k$ in the squeezing factor, using the value of $\delta_k$ determined by the Starobinsky background via \eqref{eq:delta-k-Starobinsky} and $n_{\mathrm{s}}=0.964$.}
}
	\label{fig:plot1}
\end{figure*}

%Comparison of the Planck 2018 CMB TT power spectrum data  with theoretical predictions from the base $\Lambda$CDM model, including the choices for the primordial power spectrum shown in the left panel (squeezed and power-law). The curves are computed with the  \texttt{CAMB} software \cite{Lewis:1999bs}, using the best-fit parameters reported by Planck \cite{Planck:2018vyg}.

Up to this point we have used the variable $y = - k \tau$, where $\tau$ is a generalized conformal time, and the expressions in the previous sections are evaluated around a peculiar time $\tau_k=-1/k$, as in \cite{Stewart2001, Auclair2022,Bianchi:2024qyp}. Cosmological observations probe the power spectrum in a finite window in $k$. For instance, the CMB scales observed by the Planck mission are in the range between $10^{-4}\,\mathrm{Mpc}^{-1}$ and $10^{-1}\,\mathrm{Mpc}^{-1}$\cite{Planck:2018jri}. In order to compare to observations, it is useful to determine the power spectrum fully expanded around a pivot scale $k_\ast$. The running of the power spectrum along a range of values of $k$ around a pivot scale $k_\ast$ can be obtained by evaluating the quantities at a generalized conformal time $\tau_\ast = -1/k_\ast$, so that it corresponds to a slightly modified horizon crossing condition around $y_\ast = k/k_\ast$, i.e, $k_\ast = (a_\ast H_\ast)/\tilde{c}_\ast$, as described in Appendix \ref{eq:App-B}. Using $\ln(y_k/y_\ast) = - \ln(k/k_\ast)$, we find that the phase $\delta_{k\ast}$ evaluated at the pivot scale $k_\ast$ has the form
\begin{equation}
\delta_{k\ast}=\pi\,\Big(1-\tfrac{1}{2}p_{0\ast}\Big)p_{1\ast}+\pi \,\Big(p_{2\ast}-\tfrac{1}{2}p_{1\ast}^2\Big)\ln\left(\frac{k}{k_\ast}\right)\, ,
\label{eq:delta-k-expanded}
\end{equation}
and the fully-expanded squeezed power spectrum reads
\begin{widetext}
\begin{align}
	\label{eq:Power-Squeezed-N2LO}
& \mathcal{P}_{\mathrm{sqz}} (k) = \frac{\hbar\,H_\ast^2 \cosh(2r_k)}{4\pi^2 c_\ast^3 Z_\ast} \Bigg\{p_{0\ast} + \Big[ -\Big(p_{0\ast} - \frac{\pi^2}{8}p_{1\ast}^2 \Big)\cos(\phi_k) + \frac{\pi}{2} \, p_{1\ast} \, \sin(\phi_k) \Big] \tanh(2r_k) \nonumber\\
&\; +\Big[ p_{1\ast} - \big(p_{1\ast}\cos(\phi_k) -\pi\, p_{2\ast}\, \sin(\phi_k) \big) \tanh(2r_k)  \Big]  \ln(\frac{k}{k_\ast})  +  \Big[ p_{2\ast}  - p_{2\ast} \cos(\phi_k) \tanh(2r_k)   \Big] \ln(\frac{k}{k_\ast})^2  + \order{\textrm{N3LO}} \Bigg\}  \, ,
\end{align}
\end{widetext}
where the coefficients $p_{n\ast}$ are the same as in \eqref{eq:Power-BunchDavies}, and their exact expression in terms of Hubble-flow parameters can be found in \eqref{eq:p0}, \eqref{eq:p1}, and \eqref{eq:p2}. The expressions \eqref{eq:delta-k-expanded} and \eqref{eq:Power-Squeezed-N2LO} are the main result of this paper. They capture the effect of the squeezed vacuum for either scalar or tensor modes in a general effective theory of inflation, accurately up to N2LO in the Hubble-flow expansion.

% accurately describes the variation of the power spectrum around a pivot scale $k_\ast$ for a squeezed choice of the vacuum state of the quantum field corresponding to the perturbations of interest. By comparing \eqref{eq:Power-Squeezed-N2LO} with \eqref{eq:Power-BunchDavies}, it is clear that $ \lim_{r_k\to 0} \mathcal{P}_{\mathrm{sqz}} (k)= \mathcal{P}_{\mathrm{qBD}}(k)$, which is consistent with the general property of the squeezing pre-factor $\lim_{r_k\to 0}\Upsilon_{\mathrm{N2LO}} =1$. 

\section{Primordial Features and the effect of the phase \boldmath{$\delta_k$}}
\label{sec:NumericalAnalyis}

To illustrate the effect of the nontrivial phase $\delta_k$ on the squeezed power spectrum \eqref{eq:power-delta}, we consider a simple choice of squeezed vacuum $|\mathrm{sqz}\rangle$ given by the Bogoliubov coefficients 
\begin{equation}
\label{eq:alpha-beta-Bogoliubov}
\alpha_k = 1 -\frac{k_c^2}{2k^2} -  \ii \, \frac{k_c}{k},\quad\beta_k =  -\frac{k_c^2}{2k^2}\,\,\ee^{\,2\,\ii k/k_c}. 
\end{equation}
This specific choice has been considered and studied before (see for instance \cite{Danielsson:2002kx,Danielsson:2004xw,Broy:2016zik}, and the review \cite{Chluba:2015bqa}) and can be understood as a simplified model of the effect of a pre-inflationary phase. In this model, one considers an instantaneous transition from Minkowski space to de Sitter space, i.e., $a(t) = k_c/H_0 $ for $t<0$, and $a(t) =(k_c/H_0)\,\ee^{H_0 t}$ for $t\ge 0$, where $k_c$ is a new physical scale defined as the comoving scale at the transition time $t=0$. In this situation, there are two natural choices of state, the in-vacuum given by the Minkowski state $|\mathrm{M}\rangle$ for $t<0$ and the out-vacuum given by the Bunch-Davies state $|\mathrm{BD}\rangle$ for $t\geq 0$. They correspond to the mode functions $w_M(y)=\ee^{\ii y}$ for $y< y_c$ and $w_{\mathrm{BD}}(y)$ for $y\geq y_c$, with $y_c\equiv k/k_c$. If a quantum field is prepared in the in-vacuum, the mode functions after the transition can be written as  $w_{\mathrm{sqz}}(y)=\alpha_k\, w_{\mathrm{BD}}(y)+\beta_k\, w_{\mathrm{BD}}^*(y)$, with the Bogoliubov coefficients $\alpha_k$ and $\beta_k$ determined by the matching of the mode function and its derivative at the time $y=y_c$, i.e., $w_{\mathrm{sqz}}(y_c)=w_{\mathrm{M}}(y_c)$ and $w_{\mathrm{sqz}}'(y_c)=w_{\mathrm{M}}'(y_c)$. This model provides a well-defined fiducial prescription for the Bogoliubov coefficients in terms of one single new physical scale, the comoving scale $k_c$ at the transition, and allows us to illustrate the effect of the phase shift $\delta_k$ determined in \eqref{eq:delta-k-expanded}. Assuming the squeezed state $|\mathrm{sqz}\rangle$ is defined exactly by the Bogoliubov coefficients \eqref{eq:alpha-beta-Bogoliubov}, one finds that the squeezing factor, for $k\gg k_c$, takes the form $\Upsilon_{\mathrm{sqz}}(k) = 1-(k_c^2/k^2)\,\cos(2k/k_c+\delta_k)+\order{k_c^3/k^3}$.
This expression reproduces the form of the phenomenological template \eqref{eq:template-phase}. Moreover, using the result \eqref{eq:delta-k-expanded}, the phase shift $\delta_k$ is now completely fixed by the background geometry. 

\begin{figure}[t]
	\includegraphics[width=\linewidth]{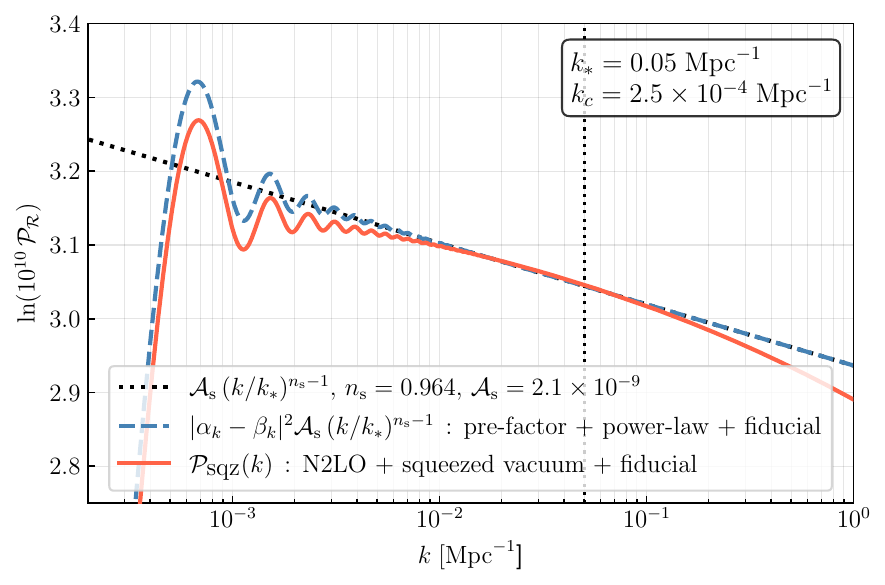}\quad
	\caption{To further illustrate the effect of the phase $\delta_k$ we compare the  N2LO result \eqref{eq:Power-Squeezed-N2LO} (solid red) with the expression with $\delta_k = 0$ artificially set to zero (dashed blue), adopting exaggerated values of the Hubble-flow parameters $\epsilon_{2H\ast} = -0.2$, $\epsilon_{2Z\ast}=-0.2$, to show a noticeable running. We compare them to the power spectrum produced by an exact power-law (dotted black), with the same Bogoliubov coefficients \eqref{eq:alpha-beta-Bogoliubov}. %(Right): We compare the squeezed power $\mathcal{P}_{\mathrm{sqz}}(k)$ with the standard expression associated to the power-law, i.e., setting artificially $\delta_k=0$ to show its effect of the power spectrum.}
    }
	\label{fig:plot2}
\end{figure}

To highlight the consequences of the choice \eqref{eq:alpha-beta-Bogoliubov}, 
we consider ---as a concrete example--- the primordial power spectrum of curvature perturbations $\mathcal{R}(k)$ in the Starobinsky model of inflation \cite{Starobinsky1979,Starobinsky1980}. As discussed in \cite{Bianchi:2024qyp}, the Hubble-flow parameters associated to the curvature perturbations are given by $\epsilon_{1H\ast} \approx 0.009$, $\epsilon_{2H\ast} \approx -0.018$, $\epsilon_{1Z\ast} \approx -0.018$, $\epsilon_{2Z\ast}\approx -0.018$, $\epsilon_{1c\ast} = 0$, and $\epsilon_{2c\ast} = 0$, which are the approximate figures at $N_\ast =55$. We will use the typical pivot scale $k_\ast = 0.05$ Mpc$^{-1}$. The new physical scale $k_c$ that characterizes the squeezed state $|\mathrm{sqz}\rangle$ via \eqref{eq:alpha-beta-Bogoliubov} is assumed here to take the fiducial value $k_c = 2.5 \times 10^{-4}$ Mpc$^{-1}$, corresponding to just-enough inflation \cite{Schwarz:2009sj,Ramirez:2011kk,Ramirez:2012gt,Cicoli:2014bja}, that is a finite duration of the inflationary phase $N_{\mathrm{infl}}\sim 60$ that is not ruled out yet by observations and leads to features in the observable window of the power spectrum. The squeezed vacuum for this model exhibits clear primordial features, as shown in the left panel of Fig.~\ref{fig:plot1}:  a power suppression below the physical scale $k_c$ and fast oscillations around $k_c$, rapidly converging to the tilted power associated to the quasi-Bunch-Davies vacuum for $k\gg k_c$. \comment{The effect of the phase $\delta_k$ is illustrated in the right panel of Fig.~\ref{fig:plot1} where we compare the squeezing parameter $|\alpha_k-\beta_k\,\mathrm{e}^{\mathrm{i}\delta_k}|^2$ with $\delta_k$ given by \eqref{eq:delta-k-expanded}, compared to the naive value of a vanishing $\delta_k$.}
The relevance of our N2LO results are further illustrated in Fig.~\ref{fig:plot2}, where we adopt an artificially enhanced set of second order Hubble-flow parameters, $\epsilon_{2H\ast}$ and $\epsilon_{2Z\ast}$. \comment{The larger values of these parameters induce a stronger dependence on $\delta_k$ that results in a running of $n_{\textrm{s}}$, which cannot be captured by a standard power-law ansatz. %In the right panel, we compare the effect of setting $\delta_k = 0$ (as often done implicitly in previous works) with the full N2LO squeezed power $\mathcal{P}_{\mathrm{sqz}}(k)$ for Starobinsky inflation. Although the difference is small in this case, the phase shift remains distinguishable, which shows that $\delta_k$ should be considered for the prediction of observable quantities.
}

%The squeezed vacuum for this model exhibits clear primordial features, as shown in the left panel of Fig.~\ref{fig:plot1}:  a power suppression below the physical scale $k_c$ and fast oscillations around $k_c$, rapidly converging to the tilted power associated to the quasi-Bunch-Davies vacuum for $k\gg k_c$. These prototypical features may have observational consequences, as illustrated in the right panel of Fig.~\ref{fig:plot1}, which shows the CMB angular power spectrum of temperature fluctuations, $\mathcal{D}_{\ell} = \ell (\ell+1)C_\ell/2\pi$, as measured by Planck \cite{Planck:2019nip}. The pre-inflationary corrections  affect the low multipoles ($\ell \lesssim 30$), where their imprint on the power spectrum becomes most evident.

Note that, as done in \cite{Bianchi:2024qyp}, in Starobinsky inflation we can express all power-law quantities in the qBD vacuum up to N2LO in terms of a single parameter, the scalar tilt $n_\mathrm{s}$ which is also one of the most accurately measured cosmological parameters, $n_\mathrm{s}-1= -0.0351\pm0.0042$ at $68\%$ C.L. \cite{Planck:2018jri}. Introducing an N2LO truncation in the parameter $|n_\mathrm{s}-1|\ll 1$, we find that the phase shift for scalar and tensor modes can be written as
\begin{align}
\delta_{k\ast}^{(\textrm{s})} &=\frac{\pi}{2}(n_\mathrm{s}-1)\,-\frac{\pi}{4}(n_\mathrm{s}-1)^2\,\ln\left(\frac{k}{k_\ast}\right)\, + \order{\textrm{N3LO}}\nonumber \\
\delta_{k\ast}^{(\textrm{t})} &=-\frac{3\pi}{16} (n_{\textrm{s}} -1)^2\, + \order{\textrm{N3LO}}.
\label{eq:delta-k-Starobinsky}
\end{align}
Besides a constant negative phase shift proportional to $n_\mathrm{s}-1$, the phase $\delta_{k\ast}^{(\textrm{s})}$ introduces a running drift $\ln(k/k_\ast)$, which is analogous to the one considered in phenomenological templates \cite{Flauger:2014ana}. Furthermore, assuming that the Bogoliubov coefficients are the same for tensor and scalar modes, the relative phases play a role in the determination of the tensor-to-scalar ratio $\mathfrak{r}_\ast^{(\textrm{sqz})} = \mathcal{P}_{\mathrm{sqz}}^{\mathrm{(t)}}  (k_\ast)/\mathcal{P}_{\mathrm{sqz}}^{\mathrm{(s)}}  (k_\ast)$. In the limit of small squeezing, $r_k\ll 1$, we have a corrected tensor-to-scalar ratio $\mathfrak{r}_\ast^{(\textrm{sqz})} \approx [1+2 \, r_{k\ast} (\delta_{k\ast}^{(\textrm{t})} - \delta_{k\ast}^{(\textrm{s})})\, \sin(\phi_{k\ast})]\, \mathfrak{r}_\ast^{(\textrm{qBD})}$. 

\newpage 

\section{Discussion}
\label{sec:Discussion}

We derived the effect of a squeezed vacuum on the primordial power spectrum. In particular, we determined the phase $\delta_k$ appearing in \eqref{eq:power-delta} and computed its fully-expanded expression \eqref{eq:delta-k-expanded} around a pivot mode $k_\ast$, together with the power spectrum \eqref{eq:Power-Squeezed-N2LO},  for scalar and tensor perturbations in a large class of effective theories of inflation \eqref{eq:Quadratic-Action-psi} characterized by a kinetic amplitude $Z_\psi(t)$ and speed of sound $c_\psi(t)$. In the case of Starobinsky inflation, we found that the phase $\delta_k$ for curvature perturbations can be written purely in terms of the scalar tilt $n_{\mathrm{s}}$, resulting in a small negative shift together with a running drift \eqref{eq:delta-k-Starobinsky}, providing precise theoretical predictions that can be used in phenomenological templates for the primordial features \eqref{eq:template-phase}. 

It would be interesting to extend the analysis of squeezed vacua at N2LO presented here to other observables, including the bispectrum \cite{Chluba:2015bqa}, and combine it with top-down proposals of new physics in a pre-inflationary phase \cite{Starobinsky:1992ts,Niemeyer:2000eh,Tanaka:2000jw,Starobinsky:2001kn,Hui:2001ce,Goldstein:2002fc,Danielsson:2002kx,Niemeyer:2002kh,Contaldi:2003zv,Martin:2003kp,Armendariz-Picon:2003knj,Bozza:2003pr,Danielsson:2004xw,Wang:2007ws,Destri:2009hn,Chen:2015gla,Broy:2016zik,Handley:2016ods,Kaloper:2018zgi,Gessey-Jones:2021yky,Letey:2022hdp,Schwarz:2009sj,Ramirez:2011kk,Ramirez:2012gt,Cicoli:2014bja,Agullo:2012sh,Fernandez-Mendez:2012poe,Fernandez-Mendez:2013jqa,Agullo:2013ai,Agullo:2015tca,deBlas:2016puz,CastelloGomar:2017kbo,Bhardwaj:2018omt,Ashtekar:2020gec,Navascues:2021mxq,Kowalczyk:2024ech,MenaMarugan:2024zcv} to constrain the effect of primordial squeezed vacua with cosmological observations.

\bigskip

\begin{acknowledgments}
We thank Miguel Fernandez, Monica Rincon-Ramirez, Javier Olmedo, Brajesh Gupt and Abhay Ashtekar for useful discussions. M.G. is supported by the \href{https://anid.cl}{Agencia Nacional de Investigación y Desarrollo} (ANID) and \href{http://www.fulbright.cl/}{ Fulbright Chile} through the Fulbright Foreign Student Program and ANID BECAS/Doctorado BIO Fulbright-ANID 56190016.~E.B. acknowledges support from the National Science Foundation, Grant No. PHY-2207851. This work was made possible through the support of the ID 62312 grant from the John Templeton Foundation, as part of the project \href{https://www.templeton.org/grant/the-quantum-information-structure-of-spacetime-qiss-second-phase}{``The Quantum Information Structure of Spacetime'' (QISS)}. The opinions expressed in this work are those of the authors and do not necessarily reflect the views of the John Templeton Foundation.
\end{acknowledgments}

% \bibliography{bib-inflation}
\newpage
%\clearpage

\appendix

% %
% \begin{center}
%     \textbf{-- End Matter --}
% \end{center}
% %

\section{Derivation of the mode equation}
\label{app:DerivationModes}

Here we summarize some of the steps in the derivation of the mode equation, following \cite{Bianchi:2024qyp}. First, note that for the original mode $u(k,t)$ satisfies the following Wronskian condition,
\begin{equation}
	\label{eq:CCR-u(t)}
	u(k,t) \dot{u}^\ast (k,t) - \dot{u}(k,t) u^\ast(k,t) = \frac{\ii \,\hbar}{2\, a(t)^3\, Z_\psi (t)},
\end{equation}
where $u^\ast$ is the complex conjugate of $u$. On the other hand, we have the mode equation
\begin{equation}
	\label{eq:EoM-u(t)}
	\ddot{u}(k,t) + (3-\epsilon_{Z1}(t)) H(t) \dot{u}(k,t) + c_\psi(t)^2\frac{k^2}{a(t)^2}  \, u(k,t) = 0\,.
\end{equation} 
Under the map $u(k,t)\to w(y)$, the above equation reads
\begin{equation}
	\label{eq:EoM-v(x)-aH}
	w''(y) +  \left[   1 + \frac{a(t)^2 H(t)^2}{k \, c_\psi^2(t)} q(t) \right] w(y)  = 0,
\end{equation}
where
\begin{align}
	q(t) & = -2 + \epsilon_{H1}(t)  + \frac{3}{2} \epsilon_{Z1}(t) + \frac{\epsilon_{1c}(t)}{2}  \nonumber \\
	&\quad -\frac{\epsilon_{1H} (t) \epsilon_{1Z} (t)}{2}  - \frac{\epsilon_{1Z} (t)^2 }{4} - \frac{ \epsilon_{1Z} (t) \epsilon_{2Z} (t)   }{2} \nonumber \\
	&\quad - \frac{ \epsilon_{1c} (t) \epsilon_{1H} (t)   }{2}-
	\frac{ \epsilon_{1c}(t) \epsilon_{2c}(t)}{2}      + \frac{\epsilon_{1c}^2(t)}{4}  .
\end{align}

The above expression is exact in $\epsilon_{1H}(t)$, $\epsilon_{1Z}(t)$, $\epsilon_{1c}(t)$, etc. Writing $t$ (or, equivalently, a generalized conformal time $\tau$) as a function of $y$ in a self-consistent way, e.g., as discussed in the Appendix A of \cite{Bianchi:2024qyp}, we find at N2LO, 
	\begin{align}
		\label{eq:y-in-terms-of-aH}
		\tau(t) &= -\frac{c_\psi(t)}{a(t)H(t)} \Big[  1 + \epsilon_{1H}(t)  - \epsilon_{1c}(t)  + \epsilon_{1H}(t)^2 \nonumber \\
        &\quad - \epsilon_{1H}(t) \epsilon_{2H}(t)  - 2 \epsilon_{1c}(t) \epsilon_{1H}(t)  \nonumber \\
        &\quad + \epsilon_{1c}(t)\epsilon_{2c}(t) + \epsilon_{1c}(t)^2  +\mathcal{O}(\epsilon^3)  \Big]\nonumber \\
		&\equiv- \frac{\, \tilde{c}_\psi (t)}{a(t)H(t)}\, .
	\end{align}

We define the variable $y \equiv - k \tau$, so we can use the above equation anytime we need $a(t)$ in terms of $y$. The next step is to write each flow parameter in terms of $y$, which can be done via the logarithmic expansion around the peculiar time $\tau_k\equiv -1/k$, i.e., around $y_k = 1$,
\begin{align}
	\label{eq:LogExpansion}
	\rho (y)
	&= \rho_k \Big[ 1 + \Big(\epsilon_{1\rho k}  + \epsilon_{1\rho k} ( \epsilon_{1H k}-\epsilon_{1c k} )  \Big) \ln\left(y\right) \nonumber \\
    &\quad+ \frac{1}{2}\,\Big( \epsilon_{1\rho k} (\epsilon_{1\rho k} + \epsilon_{2\rho k})  \Big) \ln\left(y\right)^2  + \order{\epsilon^3} \Big] \,.
\end{align}
Hence, after using the logarithmic expansion, \eqref{eq:EoM-v(x)-aH} reduces to the mode equation \eqref{eq:EOM}, with the function $g(y)$ given by
\begin{align}
& g(y)\; =\; g_{1k} + g_{2k}\, \ln(y)\,+\,\order{\epsilon^3}\,,\label{eq:gk}\\[.5em]
  &\quad g_{1k} =-\frac{3}{2} \left(  - 2 \epsilon_{1Hk} + \epsilon_{1Zk} +3 \epsilon_{1ck}\right) \nonumber \\
	&\qquad + \frac{1}{4} \Big( 27 \epsilon_{1ck}^2 - 42 \epsilon_{1ck} \epsilon_{1Hk} 
	+ 16 \epsilon_{1Hk}^2 + 12 \epsilon_{1ck} \epsilon_{1Zk} \nonumber \\
	&\qquad\qquad - 10 \epsilon_{1Hk} \epsilon_{1Zk} + \epsilon_{1Zk}^2  + 18 \epsilon_{1ck} \epsilon_{2ck} \nonumber \\
    &\qquad\qquad - 16 \epsilon_{1Hk} \epsilon_{2Hk} 
	+ 2 \epsilon_{1Zk} \epsilon_{2Zk} \Big)\,,\nonumber
	  \\
  &\quad g_{2k} = -\frac{3}{2} \left( - 2 \epsilon_{1Hk} \epsilon_{2Hk} + \epsilon_{1Zk} \epsilon_{2Zk} +3 \epsilon_{1ck} \epsilon_{2ck}\right)\,.\nonumber
\end{align}
Note that the function $\mu(y)$ can also be expanded,
\begin{align}
 & \mu(y)\; =\;\frac{Z_k \,c_k^3}{\hbar\,H_k^2}\,\Big(1+\mu_{1k}+\mu_{2k}\ln(y) \nonumber\\
 &\hspace{8em}+ \mu_{3k} \ln(y)^2\, +\order{\epsilon^3}\Big)\,, \label{eq:muk}  \\[.5em]
&\quad \mu_{1k} =   2 \epsilon_{1Hk} -2 \epsilon_{1ck}+ 3 \epsilon_{1Hk}^2 - 2 \epsilon_{1Hk} \epsilon_{2Hk} + 3 \epsilon_{1ck}^2 \nonumber\\
&\quad\qquad - 6 \epsilon_{1ck} \epsilon_{1Hk} + 2 \epsilon_{1ck} \epsilon_{2ck}\,,
 \nonumber \\[.5em]
&\quad \mu_{2k}= - 2 \epsilon_{1Hk} + \epsilon_{1Zk}+3 \epsilon_{1ck}  - 6 \epsilon_{1Hk}^2  + 3 \epsilon_{1Hk} \epsilon_{1Zk}  \nonumber \\
&\qquad \qquad + 2 \epsilon_{1Hk} \epsilon_{2Hk} - 9 \epsilon_{1ck}^2 - 2 \epsilon_{1ck} \epsilon_{2ck}- 3 \epsilon_{1ck} \epsilon_{1Zk} \nonumber \\
&\qquad \qquad + 15 \epsilon_{1ck} \epsilon_{1Hk}\,,\nonumber\\
&\quad \mu_{3k} = + 2 \epsilon_{1Hk}^2 - 2 \epsilon_{1Hk} \epsilon_{1Zk} + \frac{\epsilon_{1Zk}^2}{2}  - \epsilon_{1Hk} \epsilon_{2Hk} \nonumber \\
&\qquad \qquad + \frac{\epsilon_{1Zk} \epsilon_{2Zk}}{2}  - 6 \epsilon_{1ck} \epsilon_{1Hk}+ 3 \epsilon_{1ck} \epsilon_{1Zk}\nonumber\\
&\qquad \qquad + \frac{3 \epsilon_{1ck} \epsilon_{2ck}}{2}+\frac{9 \epsilon_{1ck}^2}{2}\,.\nonumber
\end{align}
\section{Expansion around a pivot mode}
\label{eq:App-B}
In the N2LO expansion in Hubble-flow parameters, and in the late time limit $y\to 0^+$, we find
\begin{align}
& y \,w_{\mathrm{BD}}(y)=\;+\ii\,+\,\order{y} \label{eq:limits1}\\[.5em]
	& y \,w_1(y)=-\frac{\pi}{6}\,g_{1k}\,-\,\ii \;\frac{1}{3}\,g_{1k}\,\big(C+\ln(y)\big)\,+\,\order{y}  \label{eq:limits2}\\[.5em]
	& y \,w_2(y)=\frac{\pi}{54}\Big((1+3C)g_{1k}^2-3(1-3C)g_{2k}+3g_{1k}^2\ln(y)\Big) \quad\nonumber \\
	&\qquad\qquad +\ii\;\frac{1}{216}\Big(
	(3\pi^2-48+8C+12C^2)g_{1k}^2 \nonumber \\
	&\qquad\qquad\quad\quad -3(\pi^2+8C-12C^2)g_{2k}+ \nonumber \\
	&\qquad\qquad\quad\quad +\;\big(8(1+3C)g_{1k}^2-24\,g_{2k}\big)\ln(y)
	\; \nonumber \\
	&\qquad\qquad\quad\quad + 12(g_{1k}^2-3g_{2k})\ln(y)^2\Big)+\order{y}  \label{eq:limits3}
\end{align}
where $C=\gamma_E+\ln(2)-2\simeq -0.730$.
Combining \eqref{eq:muk} with the limits \eqref{eq:limits1}-\eqref{eq:limits3}, we find the finite expression
\begin{widetext}
\begin{align}
\label{eq:qBD-Power-k}
&\mathcal{P}_{\mathrm{qBD}}^{(\mathrm{N2LO})}\;= \lim_{y\to 0^+ } \frac{\abs{y\,w_{\mathrm{qBD}}(y)}^2}{4\pi^2\,\mu(y)} =\; \frac{\hbar\, H_k^2}{4\pi^2 Z_k c_k^3} \Bigg[ 1 + \Big((2 + 3C) \, \epsilon_{1ck} - 2(1 + C) \, \epsilon_{1Hk} + C \, \epsilon_{1Zk}\Big) \qquad\nonumber \\[.5em]
&\qquad+ \frac{1}{24} \bigg(
3(-64 + 24C + 36C^2 + 9\pi^2) \, \epsilon_{1ck}^2 + 12(-6 + 4C + 4C^2 + \pi^2) \, \epsilon_{1Hk}^2 \nonumber \\
&\qquad
- 3 \, \epsilon_{1ck}\, \Big( 4(-20 + 10C + 12C^2 + 3\pi^2) \, \epsilon_{1Hk} - 2(-24 + 4C + 12C^2 + 3\pi^2) \, \epsilon_{1Zk} 
+ (16 + 16C + 12C^2 - \pi^2) \, \epsilon_{2ck} \Big)  \nonumber \\
&\qquad
- 2 \, \epsilon_{1Hk}\, \Big( 6(-8 + 2C + 4C^2 + \pi^2) \, \epsilon_{1Zk} + (-24 - 24C - 12C^2 + \pi^2) \, \epsilon_{2Hk} \Big)  \nonumber \\
&\qquad
+ \epsilon_{1Zk}\, \Big( 3(-8 + 4C^2 + \pi^2) \, \epsilon_{1Zk} + (-12C^2 + \pi^2) \, \epsilon_{2Zk} \Big)
\bigg) \;+\;\order{\epsilon^3}
  \Bigg] \,.
\end{align}
\end{widetext}
Later in the calculation, we need to expand the variables around a different time, characterized by a time $\tau_\ast = -1/k_\ast$, characterized by the pivot scale $k_\ast = a_\ast H_\ast/\tilde{c}_\ast$ This is an expansion around $y_\ast = k/k_\ast$,
	\begin{align}
	&\rho(y)
	= \rho_\ast \Big[ 1 + \Big(\epsilon_{1\rho \ast}  + \epsilon_{1\rho \ast} ( \epsilon_{1H \ast}-\epsilon_{1c \ast} )  \Big) \ln\left(\frac{y}{y_\ast}\right)  \nonumber \\
    &\qquad + \frac{1}{2}\,\Big( \epsilon_{1\rho \ast} (\epsilon_{1\rho \ast} + \epsilon_{2\rho \ast})  \Big) \ln\left(\frac{y}{y_\ast}\right)^2  + \order{\epsilon^3} \Big] \,.
\end{align}
Evaluating the above expression at $y_k$, gives the translation between quantities evaluated at the peculiar time $\tau_k = -1/k$ and the pivot time $\tau_\ast = -1/k_\ast$. Since $\ln(y/y_\ast) = -\ln(k/k_\ast)$, the expansion reads,
\begin{align}
		\label{eq:LogExpansion-k/ks}
	&\rho_k = \rho(y_k)
	= \rho_\ast \Big[ 1 -\Big(\epsilon_{1\rho \ast}  + \epsilon_{1\rho \ast} ( \epsilon_{1H \ast}-\epsilon_{1c \ast} )  \Big) \ln\left(\frac{k}{k_\ast}\right) \nonumber \\
    &\qquad + \frac{1}{2}\,\Big( \epsilon_{1\rho \ast} (\epsilon_{1\rho \ast} + \epsilon_{2\rho \ast})  \Big) \ln\left(\frac{k}{k_\ast}\right)^2  + \order{\epsilon^3} \Big] \,\\
   & \epsilon_{n\rho k} 
	\equiv \epsilon_{n\rho} \qty(\frac{y_k}{y_\ast})=\epsilon_{n\rho\ast}  - \epsilon_{n\rho\ast} \epsilon_{n+1\rho\ast}  \ln\left(\frac{k}{k_\ast}\right).
\end{align}
Direct computation including the above expansion gives,
\begin{align}
\mathcal{P}_{\mathrm{qBD}}^{(\mathrm{N2LO})} (k) &= \frac{\hbar\, H_\ast^2}{4\pi^2 Z_\ast c_\ast^3} \Bigg[  p_{0\ast} + p_{1\ast} \ln(\frac{k}{k_\ast} ) + p_{2\ast} \ln(\frac{k}{k_\ast})^2 \Bigg] ,
\end{align}
with the coefficients,
\begin{widetext}
\begin{align}
p_{0\ast} &=1\,- 2 (1 + C) \epsilon_{1H\ast} + C \epsilon_{1Z\ast}+(2 + 3 C) \epsilon_{1c\ast}  \nonumber\\
&\quad + \frac{1}{24} \Bigg( 3 (-64 + 24 C + 36 C^2 + 9 \pi^2) \epsilon_{1c\ast}^2 
+ 12 (-6 + 4 C + 4 C^2 + \pi^2) \epsilon_{1H\ast}^2 \nonumber\\
&\quad - 3 \epsilon_{1c\ast} \Big( 4 (-20 + 10 C + 12 C^2 + 3 \pi^2) \epsilon_{1H\ast} 
- 2 (-24 + 4 C + 12 C^2 + 3 \pi^2) \epsilon_{1Z\ast} 
+ (16 + 16 C + 12 C^2 - \pi^2) \epsilon_{2c\ast} \Big) \nonumber\\
&\quad - 2 \epsilon_{1H\ast} \Big( 6 (-8 + 2 C + 4 C^2 + \pi^2) \epsilon_{1Z\ast} 
+ (-24 - 24 C - 12 C^2 + \pi^2) \epsilon_{2H\ast} \Big)\nonumber \\
&\quad + \epsilon_{1Z\ast} \Big( 3 (-8 + 4 C^2 + \pi^2) \epsilon_{1Z\ast} 
+ (-12 C^2 + \pi^2) \epsilon_{2Z\ast} \Big) 
\Bigg) \label{eq:p0} \\
p_{1\ast} &=  3 \epsilon_{1c\ast} - 2 \epsilon_{1H\ast} + \epsilon_{1Z\ast} \nonumber \\
&\quad + \Bigg( (3 + 9 C) \epsilon_{1c\ast}^2 
+ (2 + 4 C) \epsilon_{1H\ast}^2 
+ \epsilon_{1c\ast} \left( - (5 + 12 C) \epsilon_{1H\ast} + \epsilon_{1Z\ast} + 6 C \epsilon_{1Z\ast} - 2 \epsilon_{2c\ast} - 3 C \epsilon_{2c\ast} \right) \nonumber \\
&\quad - \epsilon_{1H\ast} \left( \epsilon_{1Z\ast} + 4 C \epsilon_{1Z\ast} - 2 (1 + C) \epsilon_{2H\ast} \right) 
+ C \epsilon_{1Z\ast} \left( \epsilon_{1Z\ast} - \epsilon_{2Z\ast} \right) 
\Bigg) \label{eq:p1} \\
p_{2\ast} &= \frac{1}{2} \Bigg( 9 \epsilon_{1c\ast}^2 + 4 \epsilon_{1H\ast}^2 
- 4 \epsilon_{1H\ast} \epsilon_{1Z\ast} + \epsilon_{1Z\ast}^2 
- 3 \epsilon_{1c\ast} \left( 4 \epsilon_{1H\ast} - 2 \epsilon_{1Z\ast} + \epsilon_{2c\ast} \right) 
+ 2 \epsilon_{1H\ast} \epsilon_{2H\ast} - \epsilon_{1Z\ast} \epsilon_{2Z\ast} 
\Bigg) \label{eq:p2}
\end{align}
\end{widetext}

\vfill
%\twocolumngrid

%\bibliographystyle{apsrev4-1}
%%\bibliographystyle{JHEP}

%\bibliography{bib-inflation}


%spphys
\end{document}